\documentclass[final,3p,times,twocolumn]{elsarticle}

\usepackage{amssymb}
\usepackage{dblfloatfix}

\journal{Journal of Crystal Growth}

\begin{document}

\begin{frontmatter}



\title{Floating zone growth of $\alpha$-Na$_{0.90}$MnO$_2$ single crystals}


\author[1,2]{Rebecca Dally}
\author[3]{Rapha\"{e}le J. Cl\'{e}ment}
\author[4]{Robin Chisnell}
\author[2]{Stephanie Taylor}
\author[2]{Megan Butala}
\author[5,6]{Vicky Doan-Nguyen}
\author[7]{Mahalingam Balasubramanian}
\author[4]{Jeffrey W. Lynn}
\author[3]{Clare P. Grey}
\author[2]{Stephen D. Wilson}
\ead{stephendwilson@engineering.ucsb.edu}

\address[1]{Department of Physics, Boston College, Chestnut Hill, Massachusetts 02467, USA}
\address[2]{Materials Department, University of California, Santa Barbara, California 93106-5050, USA}
\address[3]{Department of Chemistry, University of Cambridge, Lensfield Road, Cambridge CB2 1EW, United Kingdom}
\address[4]{NIST Center for Neutron Research, National Institute of Standards and Technology, Gaithersburg, Maryland 20899, USA}
\address[5]{California NanoSystems Institute, University of California, Santa Barbara, Santa Barbara, CA 93106, USA}
\address[6]{Materials Research Laboratory, University of California, Santa Barbara, Santa Barbara, CA 93106, USA}
\address[7]{Advanced Photon Source, Argonne National Laboratory, 9700 S. Cass Avenue, Argonne, Illinois 60439, USA}

\begin{abstract}
Single crystal growth of $\alpha$-Na$_{x}$MnO$_2$ ($x = 0.90$) is reported via the floating zone technique. The conditions required for stable growth and intergrowth-free crystals are described along with the results of trials under alternate growth atmospheres. Chemical and structural characterizations of the resulting $\alpha$-Na$_{0.90}$MnO$_2$ crystals are performed using ICP-AES, NMR, XANES, XPS, and neutron diffraction measurements. As a layered transition metal oxide with large ionic mobility and strong correlation effects, $\alpha$-Na$_{x}$MnO$_2$ is of interest to many communities, and the implications of large volume, high purity, single crystal growth are discussed.  
\end{abstract}





\end{frontmatter}


\section{Introduction}
\label{Introduction}

Two-dimensional layered transition metal oxides of the form ABO$_2$ (A=alkali metal, B=transition metal) have drawn the attention of scientists from a variety of backgrounds due to their wide array of novel electronic and functional properties. For instance, in the realm of novel cathode materials, $\alpha$-Na$_x$MnO$_2$ ($\alpha$-NMO) with the monoclinic NaNiO$_2$ structure type is widely studied as a potential Na-based cathode platform due to its superior cycling performance and operating potential \cite{Ma01012011, MENDIBOURE1985323}.  At the same time, $\alpha$-NMO also holds interest for researchers in the area of fundamental condensed matter physics due to its underlying anisotropic triangular lattice of Mn$^{3+}$ moments and its rich electronic phase diagram accessible via deintercalation \cite{Na58MnO2}.  While small volume crystals of the $\alpha$-NMO system have been produced via hydrothermal and sealed crucible techniques \cite{Hirano, JansenHoppe}, high purity floating zone (FZ) growth of large volume crystals has remained elusive.  This is primarily due to the challenge of dual Na and Mn volatility during growth as well as two competing polymorphic forms for NMO, namely ${\alpha}$-NMO and ${\beta}$-NaMnO$_2$ (${\beta}$-NMO), with very close energetics \cite{Clement01012015, Abakumov, Billaud, ClementCM}. The successful synthesis of large volume FZ grown crystals of the form ABO$_2$, such as Na$_x$CoO$_2$ \cite{Prabhakaran, ChenNCO, Sekar2011675, Lin2007471}, have historically provided access to deeper experimental insights. The current absence of high purity, large volume FZ crystals of $\alpha$-NMO presents an impasse to the community's understanding of this system's rich phase behavior, where recent studies have been limited to polycrystalline specimens \cite{Na58MnO2, Abakumov, Giot, Zorko2008, Zorko2014, Stock}.

$\alpha$-NMO is composed of alternating layers of two-dimensional manganese oxide and sodium sheets as shown in Fig.\ 1. It crystallizes in the monoclinic $C2/m$ spacegroup with an $O3$ layering sequence where Na ions occupy octahedrally coordinated sites between the MnO$_6$ layers.  MnO$_6$ octahedra within these layers form an edge sharing triangular lattice where the Mn$^{3+}$ cations undergo a large, cooperative Jahn-Teller distortion \cite{Giot}.  This results in an anisotropic triangular lattice of manganese cations in the high-spin $S=2$ state and a ($d^4$, $t_{2g}^{3}e_{g}^{1}$) electronic configuration. In contrast, the $\beta$ polymorph is made up of alternating zigzag-like layers of Na and MO$_2$ sheets and has orthorhombic symmetry ($Pnmn$ space group). The removal of Na from the $\alpha$ polymorph provides a means of hole doping, which introduces Mn$^{4+}$ cations and locally relaxes the Jahn-Teller distortion in the manganese oxide planes. This can lead to a rich interplay between Na ion/vacancy ordering, charge ordering, magnetic correlations, and Jahn-Teller lattice distortions as the Na-site occupancy is tuned \cite{Na58MnO2}.  While studies of this interplay in $\alpha$-NMO are just beginning, detailed investigations of single crystal specimens are notably lacking.  Successful FZ crystal growth was previously harnessed to explore and gain considerable physical insight into the structurally related Na$_x$CoO$_2$ system \cite{PhysRevLett.92.197201}, suggesting a similar approach for the crystal growth of $\alpha$-NMO.  

Here we report the FZ growth of single crystals of $\alpha$-NMO with $x = 0.90$. While $\alpha$-NMO is prone to disorder from stacking faults (SF) and intergrowths of the competing polymorph ${\beta}$-NMO---both of which originate from the tendency of the structure to twin \cite{Clement01012015, Abakumov, Billaud}---by tailoring the crystal growth speed we were able to mitigate this intergrowth contamination. Specifically, $^{23}$Na nuclear magnetic resonance (NMR) measurements characterizing the degree of structural faulting demonstrate that the growth rate correlates to the relative phase fractions of $\alpha$- and $\beta$- polymorphs and the number of stacking faults within the resulting crystals. This fact, along with neutron powder and single crystal diffraction data, show that the optimized, large volume $\alpha$-NMO crystals are free of both local and long-range ${\beta}$-phase intergrowths with a good quality mosaic. Combined x-ray absorption near edge spectroscopy (XANES), x-ray photoelectron spectroscopy (XPS), and inductively coupled plasma atomic emission spectroscopy (ICP-AES) data determine the stoichiometry of crystals grown under optimal conditions to be Na$_{0.90}$MnO$_{2}$. Our work opens the $\alpha$-NMO system to new avenues of investigation via single crystal studies harnessing a variety of experimental techniques, such as neutron scattering where large volume single crystals are required.

\section{Experimental Details}
\label{Experimental Details}
\subsection{Powder Synthesis and Crystal Growth}
Starting powders were prepared from Na$_2$CO$_3$ and MnCO$_3$ powders (Alfa Aesar, Puratronic{\textregistered} 99.997$\%$ and 99.985$\%$, respectively). The powders were mixed with a 1:1 molar ratio, plus $10\%$ weight excess of Na$_2$CO$_3$ to account for sodium loss during synthesis. The mixed powder was sintered in an alumina crucible at 350 $^{\circ}$C for 15 hours, reground and then sintered at 750 $^{\circ}$C for 15 hours. The powder was then reground, formed into a rod with a diameter of 5 mm, and pressed at 50,000 psi in an isostatic press. The pressed rod was then sintered in a vertical furnace at 1000 $^{\circ}$C for 15 hours and then quenched in air. At this point in the synthesis process, the polycrystalline rod is comprised of a majority of $\beta$-NMO. We note here that quenching was used as a preventative measure to avoid decomposition of the rod into mixed phases, and the effect of alternatively slow cooling the sintered feed rod is not explored here. The polycrystalline sintered $\beta$-rod was then cut and used as both the feed rod and as a polycrystalline seed for floating zone growth in a four mirror optical floating zone furnace with 500 W halogen lamps (Crystal Systems Corp. Model FZ-T-10000-H-VI-VPO-I-HR-PC). A 4:1 ratio of Ar:O$_2$ was used to pressurize the chamber to 0.15 MPa in order to help mitigate Na volatility, and gases were flowed through the growth chamber at rates of 80 SCCM and 20 SCCM for Ar and O$_2$, respectively. Once grown and cooled, crystals were immediately transferred to an Ar-filled glovebox for storage and further analysis. 

\subsection{Inductively coupled plasma atomic emission spectroscopy (ICP-AES)}
To determine sodium and manganese concentrations in $\alpha$-NMO crystals and polycrystalline samples, ICP-AES measurements were performed in a Thermo iCap 6300. Samples for analysis were prepared by first massing the starting materials on a 0.01 mg resolution balance and then dissolving the crystals in concentrated trace metals grade hydrochloric acid (High-Purity Standards). Heat was applied via a hot water bath over a hotplate to the samples in a closed container containing HCl until no particulates could be seen and the solution became clear. This reduced the room temperature dissolution time in HCl from 4-10 days to only 1-3 hours. Upon cooling, the dissolved sodium manganese oxide was diluted with deionized water to obtain a $5\%$ HCl matrix. Instrument calibrations for Na and Mn were done using blank, low, and high PPM solutions within a $5\%$ HCl matrix, which were prepared using standard analysis grade solutions of Na (1000 $\mu$g/mL in $1\%$ HCl) and Mn (1000 $\mu$g/mL in $2\%$ HCl) from High-Purity Standards.

\subsection{$^{23}$Na solid-state NMR (ssNMR)}
$^{23}$Na ssNMR spectra were acquired at room temperature on a Bruker Advance III 200 wide-bore spectrometer (4.7 T external magnetic field) at a Larmor frequency of -53.0 MHz. All NMR experiments were performed under 60 kHz magic angle spinning (MAS) using a 1.3 mm double-resonance HX probe and a recycle delay of 30 ms. $^{23}$Na NMR data were acquired on finely ground samples of single crystal NMO. $^{23}$Na NMR chemical shifts were referenced against solid $^{23}$NaCl at 7.21 ppm. $^{23}$Na spin echo NMR spectra were acquired using a 90$^{\circ}$ radiofrequency (RF) pulse of 1.03 $\mu$s and a 180$^{\circ}$ RF pulse of 2.06 $\mu$s at 25.04 W. Transverse (T$_2$$^{\prime}$) relaxation times were obtained from an exponential fit of the decay of the signal intensity obtained as the echo delay was increased in an NMR spin echo pulse sequence.

\subsection{X-Ray Absorption Near Edge Spectroscopy (XANES)}
XANES data were taken at beamline 20-BM-B at the Advanced Photon Source at Argonne National Laboratory with an incident energy tuned to the Mn K-edge. Single crystals of NMO were finely ground and a thin, uniform layer of powder was sealed between pieces of kapton tape under an inert environment. The standards used, LiMn$_2$O$_4$ and Mn$_2$O$_3$, were prepared in a similar manner. Data were deglitched, calibrated, and normalized using the software Athena \cite{Athena}. Mn foil was used as a reference, and a simultaneous spectrum of the foil was collected in transmission mode during each run of the sample and standards. Calibrations to each data set were made by matching the absorption edge of the Mn foil to 6539 eV and then shifting the data set by that amount \cite{XANES}.

\subsection{X-ray photoelectron spectroscopy (XPS)}
Data were taken using a Kratos Axis Ultra X-ray Photoelectron Spectroscopy system with a pass energy of 40 eV and step size of 0.1 eV. Data were analyzed using the splitting of the Mn 3$s$ peak, which is a result of the exchange coupling between 3$s$ holes and 3$d$ electrons. The NMO spectrum was corrected using a Shirley background and peaks were fit to a Gaussian-Lorentzian line shape.

\subsection{Neutron diffraction measurements}
Neutron powder diffraction data were collected using the BT-1 neutron powder diffractometer at the NIST Center for Neutron Research (NCNR). A Cu(311) monochromator with a 90$^{\circ}$ take-off angle, ${\lambda}=1.5397(2)$ \AA, and in-pile collimation of 60$^{\prime}$ were used. Data were collected over the 2${\theta}$ range of 3-168$^{\circ}$ with a step size of 0.05$^{\circ}$. About 3 g of crystal from a single growth run was ground and sealed in a vanadium container of length 50 mm and diameter 9.2 mm inside a dry He-filled glovebox. A fit to the data was calculated using the Le Bail refinement \cite{LEBAIL} option in FullProf \cite{FullProf}. 

The triple-axis instrument BT-7 \cite{Lynn2012} at NCNR was used to demonstrate the mosaic of the typical crystals using a vertically focused PG(002) monochromator and an incident energy of 14.7 meV. A single crystal of ${\sim}$0.5 g was aligned in the HK0 plane using open--25$^{\prime}$--25$^{\prime}$--120$^{\prime}$ collimators placed before the monochromator, before the sample, after the sample, and before the detector, respectively. Uncertainties where indicated represent one standard deviation.

\section{Results and Discussion}
\label{Results and Discussion}

A number of varying growth speeds and translation rates were attempted with key results summarized in Table 1. The optimal growth conditions for phase pure ${\alpha}$-NMO were found to be a 20 mm/hr mirror translation rate, 2 mm/hr feed rod translation rate, 30 rpm seed rod rotation, and 20 rpm feed rod rotation. Under these conditions, attempts to seed from a previously grown ${\alpha}$-NMO crystal were unsuccessful, likely due to substantial decomposition (\textit{i.e.} Na loss) of the seed crystal during the initial heating process. However, seeding from a polycrystalline rod was able to repeatedly nucleate a single grain crystal after ${\approx}$4 cm of growth. Facets form readily after the start of growth leading to the formation of a single domain within 4 cm, negating the need for a seed crystal. Specifically, a stable molten zone which leads to ${\alpha}$-phase growth with minimal stacking faults (the determination of which is discussed later) was achieved by starting mirror translation at 50 mm/hr from the initial polycrystalline seed and then stepping it down gradually toward 20 mm/hr, where steady state growth was performed. We note that our attempts at seeding growth at this eventual lower growth rate failed to maintain a stable molten zone. This is illustrated in Fig.\ 2, where the boule's cross section becomes more elliptical in shape with flat facets forming perpendicular to the direction of seed translation at the point where a stable molten zone was achieved at 20 mm/hr. These perpendicular facets are oriented along the (-101) lattice plane, and the crystal growth direction is along the short $b$-axis. Substantial evaporation of both Na and Mn occurred during FZ growth, and depositions composed of a mixture of Na and Mn oxides built up on the inner quartz walls of the growth chamber. We found it necessary to increase the power of the lamps slightly over the course of growth ($\approx$1-2\%) to compensate for the decreasing transparency of the tube.

ICP-AES analysis of samples grown at a rate of 20 mm/hr indicated a Na:Mn ratio of 0.90:1. We note here that the absolute values of the measured Na and Mn content in our samples also matched the reported ratios ($i.e.$ the measured Mn content was stoichiometric within experimental error), and that various sections of the rod were tested to check for consistency. The portions of the crystals grown under the stepped down growth rates showed uniform Na content across each crystal.  The results are summarized as Na:Mn ratios in Table 1 for a series of representative samples as well as the starting polycrystalline feed rod. As a further step, the relative fraction of Mn$^{3+}$ versus Mn$^{4+}$ was probed via XANES measurements, which when combined with ICP-AES results are capable of resolving substantial oxygen non-stoichiometry. XANES data (Fig.\ 3(a)) on a 20 mm/hr grown crystal show the white-line peak position close to that of the Mn$_2$O$_3$ standard with an oxidation state of Mn$^{3+}$. There is, however, a resolvable shift of the NMO spectrum toward the LiMn$_2$O$_4$ standard with an average valence of Mn$^{3.5+}$, consistent with the known Na deficiency of the sample. 

The pre-edge region of the XANES spectra is associated with transitions from the 1$s$ states to the split $t_{2g}$ and $e_{g}$ $d$-orbitals, resulting in varying peak shapes for Mn$^{4+}$ and Mn$^{3+}$ cations in varying local environments \cite{Chalmin2009}. A double peak structure in this energy range is conventionally indicative of Mn$^{4+}$, and a single broad peak is associated with Jahn-Teller distorted MnO$_{6}$ octahedra \cite{Sassin_XANES, Kwon1999510}. The double peak is resolvable in the Mn$^{3.5+}$ LiMn$_2$O$_4$ standard as shown in the inset of Fig.\ 3(a), but is not in the ${\alpha}$-NMO sample, again, indicating the majority of manganese in the sample is Mn$^{3+}$. We therefore performed XPS measurements in order to gain a more quantitative understanding of the manganese valence state. There exists a linear relationship between the manganese oxidation state and the exchange splitting of the Mn 3$s$ peak, ${\Delta}E_{3s}$ where $V_{Mn}=7.875-0.893{\Delta}E_{3s}$ \cite{SongXPS, LiXPS}. Using this relation to evaluate the Mn valence for a typical ${\alpha}$-Na$_x$MnO$_{2{\pm}{\delta}}$ crystal grown at 20 mm/hr, the data shown Fig.\ 3(b) reveal ${\Delta}E_{3s}=5.38$ eV, which corresponds to an average Mn valence of $+3.07$ $\pm$ $0.04$. The combined XPS and ICP-AES analysis of optimal ${\alpha}$-phase crystals determines the oxygen to be stoichiometric within error.

The lattice structure of ${\alpha}$-NMO crystals was verified by cutting a crystal from the end of the growth boule, crushing the crystal into powder, and then performing neutron powder diffraction.  Neutron powder data collected at 300 K are shown in Fig.\ 4(a) and can be fully indexed to the reported ${\alpha}$-NMO space group, $C2/m$, with Le Bail refined lattice parameters $a=5.6672$ {\AA} $\pm$ $0.0003$ {\AA}, $b=2.8606$ {\AA} $\pm$ $0.0001$ {\AA}, $c=5.8007$ {\AA} $\pm$ $0.0003$ {\AA}, and ${\beta}=113.143^{\circ}$ $\pm$ $0.003^{\circ}$. Separate single crystal neutron diffraction measurements on a crystal observed only a single grain with an observed full-with-at-half-maximum (FWHM) of $0.41^{\circ}$ $\pm$ $0.01^{\circ}$ as plotted in Fig.\ 4(b), which after correction for the instrumental resolution indicates a mosaic spread of $0.35^{\circ}$.  Together these measurements establish the $long$-$range$ ordered lattice structure of crystals grown under optimal conditions to phase pure ${\alpha}$-NMO single crystals; however, they are not directly sensitive to $local$ intergrowths of ${\beta}$-NMO which may arise as a series of stacking faults within the $O3$ layered structure. 

To investigate the presence of local intergrowths of ${\beta}$-NMO and stacking faults within the lattice of ${\alpha}$-NMO crystals, $^{23}$Na solid-state NMR (ssNMR) data were collected with results plotted in Fig. 5. If a number of $^{23}$Na resonant frequencies are resolved in the NMR spectra collected on crushed ${\alpha}$-NMO crystals, it suggests the presence of multiple chemical environments reflective of the formation of stacking faults (twin planes) between nanodomains of the ${\alpha}$ and ${\beta}$ polymorphs of NMO. While structural intergrowths and the formation of stacking faults between the $\alpha$- and $\beta$-polymorphs of NMO have been reported previously \cite{Clement01012015, Abakumov, Billaud}, quantifying their relative abundance across a macroscopic sample presents a challenge. Recently, a $^{23}$Na NMR study of ${\beta}$-NMO identified three resonances with isotropic shifts of ca.\ 750, 530 and 320 ppm and assigned them to Na nuclei in  ${\alpha}$-NMO domains, in ${\beta}$-NMO domains, and Na atoms in the direct vicinity of localized stacking faults, respectively \cite{Billaud}. This assignment was confirmed by recent first principles calculations of Na NMR parameters in various NMO structures containing twin planes between ${\alpha}$- and ${\beta}$-type structural domains \cite{ClementCM}. In the present work, we use these assignments to quantify the proportion of Na nuclei in these three different regions within our FZ grown NMO crystals. 

Relative fractions of Na site occupations were determined by integration of spin echo spectra shown in Fig.\ 5, and contributions from individual Na sites were scaled by a transverse relaxation factor accounting for the loss of NMR signal intensity over the signal acquisition time. Fig.\ 5(b) was collected on a crystal grown at 20 mm/hr and indicates the dominance of a single Na crystallographic environment where ca.\ $96\%$ of Na in the sample resides in an ${\alpha}$-NMO environment with a small percentage (ca.\ $4\%$) of Na near stacking faults. A nearly negligible fraction ($<1\%$) of Na in ${\beta}$-like environments indicates local ${\beta}$-NMO regions. This demonstrates that the lattice structure of crystals grown at the lowest rate of 20 mm/hr is largely free of faulting and that the local structure is consistent with the long-range ${\alpha}$-NMO crystal structure determined via neutron diffraction. The fraction of Na$^{+}$ ions in an ${\alpha}$-like environment can be further broken down into Na$^{+}$ ions close to a Mn$^{4+}$ ion (ca.\ 6\% of all Na) and into Na$^{+}$ ions surrounded by Mn$^{3+}$ ions only (ca.\ 90\%  of all Na).  The former environment is indicated by a Na resonance at 950 ppm (see small peak on the left of the alpha peak in Fig.\ 5(b)) which determines the proportion of Mn$^{4+}$ ions/Na vacancies to be in relatively good agreement with the total Na content obtained with XPS/ICP-AES. At this time, control over the sodium content is limited to $x$ = 0.90 for quality, phase-pure samples. 

Having established that phase pure ${\alpha}$-NMO single crystals can be grown via FZ, one further question explored was the degree through which the NMO polymorphs can be selected via the crystal growth rate. While the ${\beta}$-phase of NMO is nominally the higher temperature structure \cite{Ma01012011, MENDIBOURE1985323, Velikokhatnyi}, ${\beta}$-NMO is known to persist at ambient conditions through quenching the system into a metastable state \cite{Ma01012011, Abakumov, Billaud}. As a result, an increased crystal pull rate can potentially be harnessed to increase the relative phase fraction of ${\beta}$-NMO within Na$_x$MnO$_2$ crystals. To investigate this, crystals were grown under identical conditions as the optimal ${\alpha}$-Na$_{0.90}$MnO$_{2}$ crystals discussed previously with the exception of an increase in the sustained mirror translation rate to 50 mm/hr. The $^{23}$Na ssNMR spectrum collected on crystals grown under this increased rate is plotted in Fig.\ 5(a) and is dominated by the characteristic signal from Na ions in ${\beta}$-NMO type local environments; specifically, the majority phase fraction of the more rapidly grown sample is $66\%$ ${\beta}$-NMO, which far exceeds the relative fractions of $15\%$ ${\alpha}$-NMO and $19\%$ of Na near locally faulted regions. This demonstrates that the dominant growth mode has switched to the metastable ${\beta}$-NMO polymorph.  We note that this local phase mixture between ${\alpha}$-NMO, ${\beta}$-NMO, and faulted regions is consistent with the composition of powders whose long-range lattice structure is ${\beta}$-NMO---a lattice known to be highly defect prone \cite{Billaud, ClementCM} and intermixed with regions of the competing ${\alpha}$-phase. Due to the defect prone lattice of ${\beta}$-NMO, it is currently unclear whether even higher growth rates ($>$50 mm/hr) would result in more locally phase pure ${\beta}$-NMO crystals.

\section{Conclusions}
\label{Conclusions}
Floating zone crystal growth in an optical image furnace was utilized to produce large volume, single crystals of ${\alpha}$-phase Na$_x$MnO$_2$ with minimal stacking faults. ICP-AES, XANES, and XPS measurements determined that crystals grown via the parameters reported here possess a $10\%$ Na deficiency and a final stoichiometry of Na$_{0.90}$MnO$_{2}$. Further characterization of crystals grown at slower growth rates via combined neutron diffraction and ssNMR studies determined both the long-range and local structure of these crystals to be single-phase ${\alpha}$-NMO. By varying the crystal growth rate ($i.e.$ the mirror translation rate), the mixture of polymorphs present in Na$_{0.90}$MnO$_{2}$ crystals can be selected/tuned---a finding of potential interest for the creation of tailored cathode materials with a tunable intermixture of ${\alpha}$- and ${\beta}$- phases. Furthermore, the large volume growth of high purity ${\alpha}$-NMO crystals opens the compound to detailed exploration via a new array of probes such as single crystal neutron scattering and single crystal muon spin relaxation.

\section*{Acknowledgements}
SDW gratefully acknowledges support from the Hellman Foundation, and SDW and RD acknowledge support from ARO Award W911NF-16-1-0361. VDN is supported by the University of California President's Postdoctoral Fellowship and the University of California, Santa Barbara California NanoSystems Institute Elings Prize Fellowship. This work was partially supported by the Assistant Secretary for Energy Efficiency and Renewable Energy, Office of Vehicle Technologies of the U.S. Department of Energy under Contract No.\ DE-AC02-05CH11231, under the Batteries for Advanced Transportation Technologies (BATT) Program subcontract No.\ 7057154 (RJC and CPG). CPG and RJC thank the EU ERC for an Advanced Fellowship for CPG. The MRL Shared Experimental Facilities are supported by the MRSEC Program of the NSF under Award No. DMR 1121053; a member of the NSF-funded Materials Research Facilities Network (www.mrfn.org). This research used resources of the Advanced Photon Source, a U.S. Department of Energy (DOE) Office of Science User Facility operated for the DOE Office of Science by Argonne National Laboratory under Contract No.\ DE-AC02-06CH11357. Sector 20 operations are supported by the US Department of Energy and the Canadian Light Source. The identification of any commercial product or trade name does not imply endorsement or recommendation by the National Institute of Standards and Technology. 




\bibliographystyle{elsarticle-num} 
\bibliography{BibTexfile}

\begin{thebibliography}{10}
\expandafter\ifx\csname url\endcsname\relax
  \def\url#1{\texttt{#1}}\fi
\expandafter\ifx\csname urlprefix\endcsname\relax\def\urlprefix{URL }\fi
\expandafter\ifx\csname href\endcsname\relax
  \def\href#1#2{#2} \def\path#1{#1}\fi

\bibitem{Ma01012011}
X.~Ma, H.~Chen, G.~Ceder, {Electrochemical Properties of Monoclinic
  NaMnO$_{2}$}, Journal of The Electrochemical Society 158~(12) (2011)
  A1307--A1312.
\newblock \href
  {http://arxiv.org/abs/http://jes.ecsdl.org/content/158/12/A1307.full.pdf+html}
  {\path{arXiv:http://jes.ecsdl.org/content/158/12/A1307.full.pdf+html}}, \href
  {http://dx.doi.org/10.1149/2.035112jes} {\path{doi:10.1149/2.035112jes}}.

\bibitem{MENDIBOURE1985323}
A.~Mendiboure, C.~Delmas, P.~Hagenmuller, {Electrochemical intercalation and
  deintercalation of Na$_x$MnO$_2$ bronzes}, Journal of Solid State Chemistry
  57~(3) (1985) 323 -- 331.
\newblock \href
  {http://dx.doi.org/http://dx.doi.org/10.1016/0022-4596(85)90194-X}
  {\path{doi:http://dx.doi.org/10.1016/0022-4596(85)90194-X}}.

\bibitem{Na58MnO2}
X.~Li, X.~Ma, D.~Su, L.~Liu, R.~Chisnell, S.~P. Ong, H.~Chen, A.~Toumar, J.-c.
  Idrobo, Y.~Lei, J.~Bai, F.~Wang, J.~W. Lynn, Y.~S. Lee, G.~Ceder, {Direct
  visualization of the Jahn-Teller effect coupled to Na ordering in
  Na$_{5/8}$MnO$_{2}$}, Nature Materials 13 (2014) 586--92.
\newblock \href {http://dx.doi.org/10.1038/nmat3964}
  {\path{doi:10.1038/nmat3964}}.

\bibitem{Hirano}
S.-I. Hirano, R.~Narita, S.~Naka, {Hydrothermal synthesis and properties of
  Na$_{x}$MnO$_{2}$ crystals}, Journal of Crystal Growth 54 (1981) 595--599.
\newblock \href {http://dx.doi.org/10.1016/0022-0248(81)90520-0}
  {\path{doi:10.1016/0022-0248(81)90520-0}}.

\bibitem{JansenHoppe}
M.~Jansen, R.~Hoppe, {Zur Kenntnis der NaCl-Strukturfamilie Die
  Kristallstruktur von NaMnO$_{2}$}, Zeitschrift f\"{u}r anorganische und
  allgemeine Chemie 399~(2) (1973) 163--169.
\newblock \href {http://dx.doi.org/10.1002/zaac.19733990204}
  {\path{doi:10.1002/zaac.19733990204}}.

\bibitem{Clement01012015}
R.~J. Cl\'{e}ment, P.~G. Bruce, C.~P. Grey, {Review---Manganese-Based P2-Type
  Transition Metal Oxides as Sodium-Ion Battery Cathode Materials}, Journal of
  The Electrochemical Society 162~(14) (2015) A2589--A2604.
\newblock \href
  {http://arxiv.org/abs/http://jes.ecsdl.org/content/162/14/A2589.full.pdf+html}
  {\path{arXiv:http://jes.ecsdl.org/content/162/14/A2589.full.pdf+html}}, \href
  {http://dx.doi.org/10.1149/2.0201514jes} {\path{doi:10.1149/2.0201514jes}}.

\bibitem{Abakumov}
A.~M. Abakumov, A.~A. Tsirlin, I.~Bakaimi, G.~V. Tendeloo, A.~Lappas, {Multiple
  Twinning As a Structure Directing Mechanism in Layered Rock-Salt-Type Oxides:
  NaMnO$_2$ Polymorphism, Redox Potentials, and Magnetism}, Chemistry of
  Materials 26~(10) (2014) 3306--3315.
\newblock \href {http://dx.doi.org/10.1021/cm5011696}
  {\path{doi:10.1021/cm5011696}}.

\bibitem{Billaud}
J.~Billaud, R.~J. Cl\'{e}ment, A.~R. Armstrong, J.~Canales-V\'{a}zquez,
  P.~Rozier, C.~P. Grey, P.~G. Bruce, {${\beta}$-NaMnO$_2$: A High-Performance
  Cathode for Sodium-Ion Batteries}, Journal of the American Chemical Society
  136~(49) (2014) 17243--17248, pMID: 25397400.
\newblock \href {http://dx.doi.org/10.1021/ja509704t}
  {\path{doi:10.1021/ja509704t}}.

\bibitem{ClementCM}
R.~J. Cl\'{e}ment, D.~S. Middlemiss, I.~D. Seymour, A.~J. Ilott, C.~P. Grey,
  Insights into the nature and evolution upon electrochemical cycling of planar
  defects in the ${\beta}$-namno$_{2}$ na-ion battery cathode: An nmr and
  first-principles density functional theory approach, Chemistry of Materials
  28~(22) (2016) 8228--8239.
\newblock \href {http://dx.doi.org/10.1021/acs.chemmater.6b03074}
  {\path{doi:10.1021/acs.chemmater.6b03074}}.

\bibitem{Prabhakaran}
D.~Prabhakaran, A.~Boothroyd, R.~Coldea, N.~Charnley, {Crystal growth of
  Na$_x$CoO$_2$ under different atmospheres}, Journal of Crystal Growth
  271(1-2) (2004) 74 -- 80, publisher: Elsevier Science.
\newblock \href {http://dx.doi.org/10.1016/j.jcrysgro.2004.07.033}
  {\path{doi:10.1016/j.jcrysgro.2004.07.033}}.

\bibitem{ChenNCO}
D.~P. Chen, H.~C. Chen, A.~Maljuk, A.~Kulakov, H.~Zhang, P.~Lemmens, C.~T. Lin,
  {Single-crystal growth and investigation of Na$_x$CoO$_2$ and
  Na$_x$CoO$_{2}{\cdot}y$H$_2$O}, Phys. Rev. B 70 (2004) 024506.
\newblock \href {http://dx.doi.org/10.1103/PhysRevB.70.024506}
  {\path{doi:10.1103/PhysRevB.70.024506}}.

\bibitem{Sekar2011675}
{C. Sekar and S. Paulraj and P. Kanchana and B. Sch\"{u}pp-Niewa and R.
  Klingeler and G. Krabbes and B. B\"{u}chner}, {Effect of rotation of feed and
  seed rods on the quality of Na$_{0.75}$CoO$_2$ single crystal grown by
  traveling solvent floating zone method}, Materials Research Bulletin 46~(5)
  (2011) 675 -- 681.
\newblock \href
  {http://dx.doi.org/http://dx.doi.org/10.1016/j.materresbull.2011.01.027}
  {\path{doi:http://dx.doi.org/10.1016/j.materresbull.2011.01.027}}.

\bibitem{Lin2007471}
C.~Lin, D.~Chen, J.~Peng, P.~Zhang, {Growth and characterization of high
  quality single crystals of Na$_x$CoO$_2$}, Physica C: Superconductivity and
  its Applications 460–462, Part 1 (2007) 471 -- 472, proceedings of the 8th
  International Conference on Materials and Mechanisms of Superconductivity and
  High Temperature SuperconductorsM2S-HTSC \{VIII\}.
\newblock \href
  {http://dx.doi.org/http://dx.doi.org/10.1016/j.physc.2007.03.176}
  {\path{doi:http://dx.doi.org/10.1016/j.physc.2007.03.176}}.

\bibitem{Giot}
M.~Giot, L.~C. Chapon, J.~Androulakis, M.~A. Green, P.~G. Radaelli, A.~Lappas,
  {Magnetoelastic Coupling and Symmetry Breaking in the Frustrated
  Antiferromagnet ${\alpha}$-NaMnO$_{2}$}, Phys. Rev. Lett. 99 (2007) 247211.
\newblock \href {http://dx.doi.org/10.1103/PhysRevLett.99.247211}
  {\path{doi:10.1103/PhysRevLett.99.247211}}.

\bibitem{Zorko2008}
A.~Zorko, S.~{El Shawish}, D.~Ar{\v{c}}on, Z.~Jagli{\v{c}}i{\'{c}}, A.~Lappas,
  H.~{Van Tol}, L.~C. Brunel, {Magnetic interactions in ${\alpha}$-NaMnO$_2$:
  Quantum spin-2 system on a spatially anisotropic two-dimensional triangular
  lattice}, Physical Review B - Condensed Matter and Materials Physics 77~(2)
  (2008) 1--7.
\newblock \href {http://dx.doi.org/10.1103/PhysRevB.77.024412}
  {\path{doi:10.1103/PhysRevB.77.024412}}.

\bibitem{Zorko2014}
A.~Zorko, O.~Adamopoulos, M.~Komelj, D.~Ar{\v{c}}on, A.~Lappas,
  {Frustration-induced nanometre-scale inhomogeneity in a triangular
  antiferromagnet}, Nature Communications 5 (2014) 3222.
\newblock \href {http://dx.doi.org/10.1038/ncomms4222}
  {\path{doi:10.1038/ncomms4222}}.

\bibitem{Stock}
C.~Stock, L.~C. Chapon, O.~Adamopoulos, A.~Lappas, M.~Giot, J.~W. Taylor, M.~A.
  Green, C.~M. Brown, P.~G. Radaelli, {One-Dimensional Magnetic Fluctuations in
  the Spin-2 Triangular Lattice ${\alpha}$-NaMnO$_2$}, Phys. Rev. Lett. 103
  (2009) 077202.
\newblock \href {http://dx.doi.org/10.1103/PhysRevLett.103.077202}
  {\path{doi:10.1103/PhysRevLett.103.077202}}.

\bibitem{PhysRevLett.92.197201}
A.~T. Boothroyd, R.~Coldea, D.~A. Tennant, D.~Prabhakaran, L.~M. Helme, C.~D.
  Frost, {Ferromagnetic In-Plane Spin Fluctuations in Na$_{x}$CoO$_{2}$
  Observed by Neutron Inelastic Scattering}, Phys. Rev. Lett. 92 (2004) 197201.
\newblock \href {http://dx.doi.org/10.1103/PhysRevLett.92.197201}
  {\path{doi:10.1103/PhysRevLett.92.197201}}.

\bibitem{Athena}
B.~Ravel, M.~Newville, {ATHENA, ARTEMIS, HEPHAESTUS: data analysis for X-ray
  absorption spectroscopy using IFEFFIT}, Journal of Synchrotron Radiation
  12~(4) (2005) 537--541.
\newblock \href {http://dx.doi.org/10.1107/S0909049505012719}
  {\path{doi:10.1107/S0909049505012719}}.

\bibitem{XANES}
{Kraft, S. and St\"{u}mpel, J. and Becker, P. and Kuetgens, U.}, {High
  resolution x-ray absorption spectroscopy with absolute energy calibration for
  the determination of absorption edge energies}, Review of Scientific
  Instruments 67~(3) (1996) 681--687.
\newblock \href {http://dx.doi.org/http://dx.doi.org/10.1063/1.1146657}
  {\path{doi:http://dx.doi.org/10.1063/1.1146657}}.

\bibitem{LEBAIL}
A.~L. Bail, H.~Duroy, J.~Fourquet, {Ab-initio structure determination of
  LiSbWO$_6$ by X-ray powder diffraction}, Materials Research Bulletin 23~(3)
  (1988) 447 -- 452.
\newblock \href
  {http://dx.doi.org/http://dx.doi.org/10.1016/0025-5408(88)90019-0}
  {\path{doi:http://dx.doi.org/10.1016/0025-5408(88)90019-0}}.

\bibitem{FullProf}
J.~Rodriguez-Carvajal, {FullProf: A Program for Rietveld Refinement and Profile
  Matching Analysis of Complex Powder Diffraction Patterns (ILL, unpublished)}.

\bibitem{Lynn2012}
J.~Lynn, Y.~Chen, S.~Chang, Y.~Zhao, S.~Chi, W.~Ratcliff-II, B.~G. Ueland,
  R.~W. Erwin, {Double focusing thermal triple axis spectrometer at the NCNR},
  J. Research of the National Institute of Standards and Technology 117 (2012)
  61--79.
\newblock \href {http://dx.doi.org/10.6028/jres.117.002}
  {\path{doi:10.6028/jres.117.002}}.

\bibitem{Chalmin2009}
E.~Chalmin, F.~Farges, G.~E. Brown, {A pre-edge analysis of Mn K-edge XANES
  spectra to help determine the speciation of manganese in minerals and
  glasses}, Contributions to Mineralogy and Petrology 157~(1) (2009) 111--126.
\newblock \href {http://dx.doi.org/10.1007/s00410-008-0323-z}
  {\path{doi:10.1007/s00410-008-0323-z}}.

\bibitem{Sassin_XANES}
M.~B. Sassin, S.~G. Greenbaum, P.~E. Stallworth, A.~N. Mansour, B.~P. Hahn,
  K.~A. Pettigrew, D.~R. Rolison, J.~W. Long, {Achieving electrochemical
  capacitor functionality from nanoscale LiMn$_{2}$O$_{4}$ coatings on 3-D
  carbon nanoarchitectures}, J. Mater. Chem. A 1 (2013) 2431--2440.
\newblock \href {http://dx.doi.org/10.1039/C2TA00937D}
  {\path{doi:10.1039/C2TA00937D}}.

\bibitem{Kwon1999510}
O.-S. Kwon, M.-S. Kim, K.-B. Kim, {A study on the effect of lithium
  insertion–extraction on the local structure of lithium manganese oxides
  using X-ray absorption spectroscopy}, Journal of Power Sources 81–82 (1999)
  510 -- 516.
\newblock \href
  {http://dx.doi.org/http://dx.doi.org/10.1016/S0378-7753(99)00210-4}
  {\path{doi:http://dx.doi.org/10.1016/S0378-7753(99)00210-4}}.

\bibitem{SongXPS}
J.~Song, J.~Gim, S.~Kim, J.~Kang, V.~Mathew, D.~Ahn, J.~Kim, {A Sodium
  Manganese Oxide Cathode by Facile Reduction for Sodium Batteries},
  Chemistry--An Asian Journal 9~(6) (2014) 1550--1556.
\newblock \href {http://dx.doi.org/10.1002/asia.201301510}
  {\path{doi:10.1002/asia.201301510}}.

\bibitem{LiXPS}
J.-Y. Li, X.-L. Wu, X.-H. Zhang, H.-Y. Lu, G.~Wang, J.-Z. Guo, F.~Wan, R.-S.
  Wang, {Romanechite-structured Na$_{0.31}$MnO$_{1.9}$ nanofibers as
  high-performance cathode material for a sodium-ion battery}, Chem. Commun. 51
  (2015) 14848--14851.
\newblock \href {http://dx.doi.org/10.1039/C5CC05739F}
  {\path{doi:10.1039/C5CC05739F}}.

\bibitem{Velikokhatnyi}
O.~I. Velikokhatnyi, C.-C. Chang, P.~N. Kumta, {Phase Stability and Electronic
  Structure of NaMnO$_2$}, Journal of The Electrochemical Society 150~(9)
  (2003) A1262--A1266.
\newblock \href
  {http://arxiv.org/abs/http://jes.ecsdl.org/content/150/9/A1262.full.pdf+html}
  {\path{arXiv:http://jes.ecsdl.org/content/150/9/A1262.full.pdf+html}}, \href
  {http://dx.doi.org/10.1149/1.1600464} {\path{doi:10.1149/1.1600464}}.

\end{thebibliography}

%
%
%
%

\begin{table*}[!h]
 \begin{tabular}{||c c c c c c||} 
 \hline
 Sample & Na:Mn (molar ratio) & ${\alpha}$/SF/${\beta}$ ($\%$) & Growth Rate & Atmosphere & Pressure \\ [0.5ex] 
 \hline\hline
 starting powders & 1:1 + 10 wt$\%$ excess Na$_2$CO$_3$ & -- & n/a & n/a & n/a \\ 
 \hline
 ${\beta}$-rod & 1.04:1 & 6 / 37 / 57 & n/a & ambient & ambient \\
 \hline
 crystal 1 & 0.90:1 & 96 / 4 / $<$1 & 20 mm/hr & 4:1 Ar/O$_2$ & 0.15 MPa \\
 \hline
 crystal 2 & 0.94:1 & 15 / 19 / 66 & 50 mm/hr & 4:1 Ar/O$_2$ & 0.15 MPa \\
 \hline
 crystal 3 & 0.75:1 & -- & 20 mm/hr & 1:1 Ar/O$_2$ & 0.15 MPa \\
 \hline
\end{tabular}
\caption{Summary of growth trials using varying crystal pull rates and growth environments.  Compositional analyses of crystals grown under each condition as well as the polycrystalline feed material are also summarized.}
\end{table*}

\begin{figure}[!p]
\includegraphics[scale=0.25]{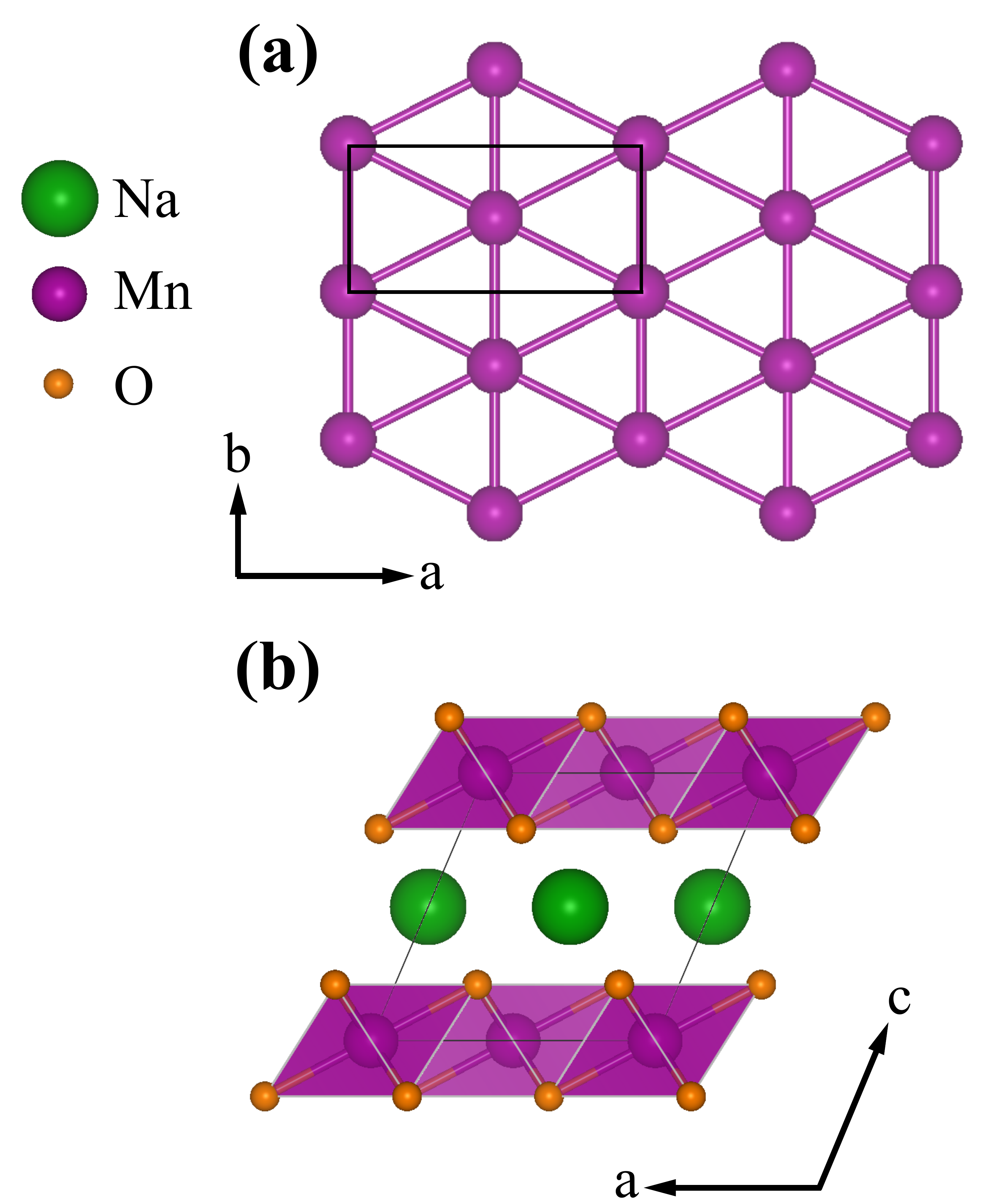}
\caption{The structure of monoclinic $\alpha$-NaMnO$_2$ is illustrated via projections of the (a) $ab$-plane and the (b) $ac$-plane, with the unit cell outlined in black for both figures. Panel (a) shows just a single plane of the manganese cations and panel (b) shows the MnO$_6$ polyhedra as shaded purple regions, oxygen atoms as orange spheres, and Na atoms as green spheres. }
\end{figure}

\begin{figure}[!p]
\includegraphics[scale=0.47]{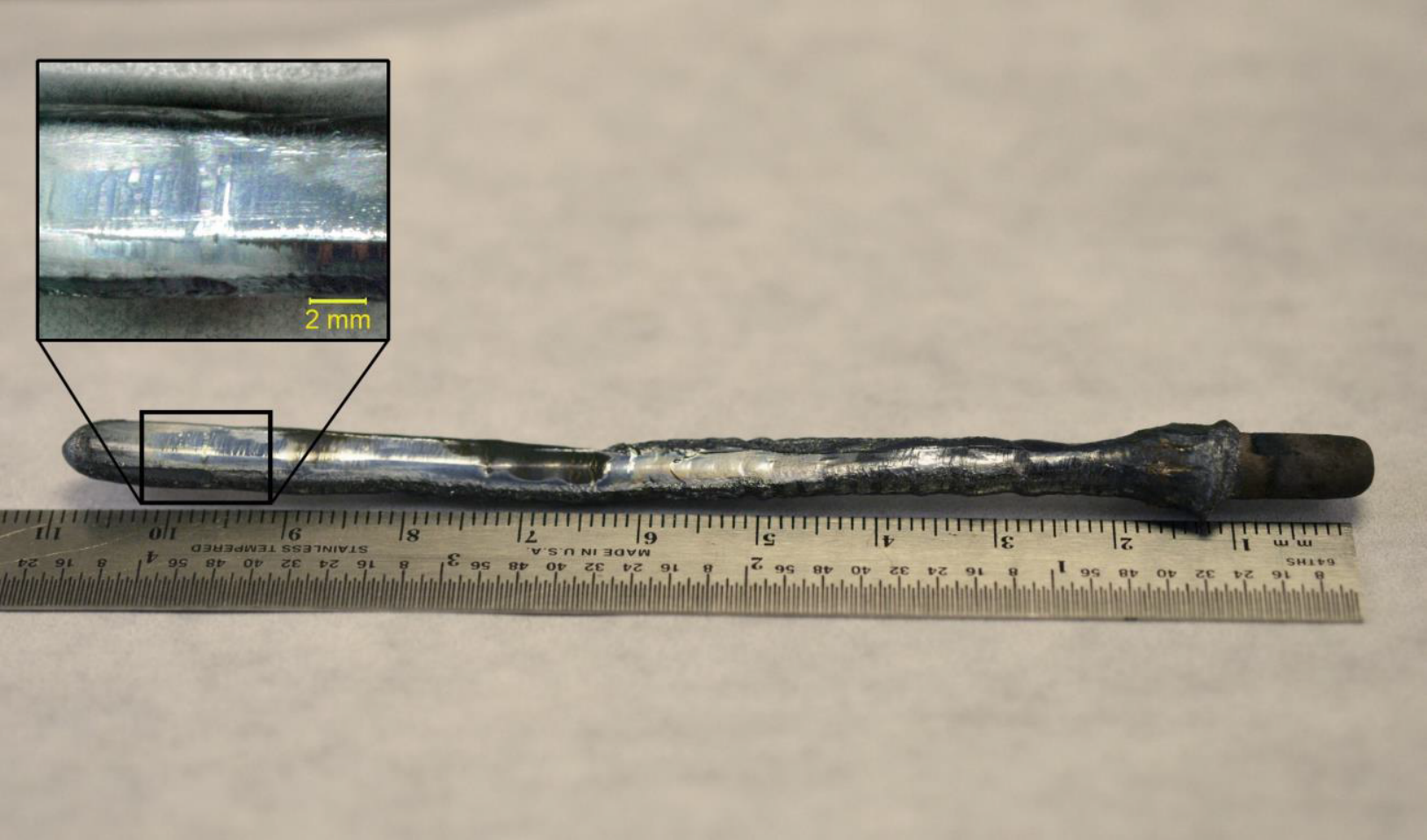}
\caption{As grown $\alpha$-Na$_{0.90}$MnO$_2$ crystal. Stable crystal growth begins after ${\approx}$4 cm of translation, where there is a visible change in shape. The inset shows a close-up of the flat (-101) facet that was formed during stable growth.}
\end{figure}

\begin{figure}[!p]
\includegraphics[scale=.6]{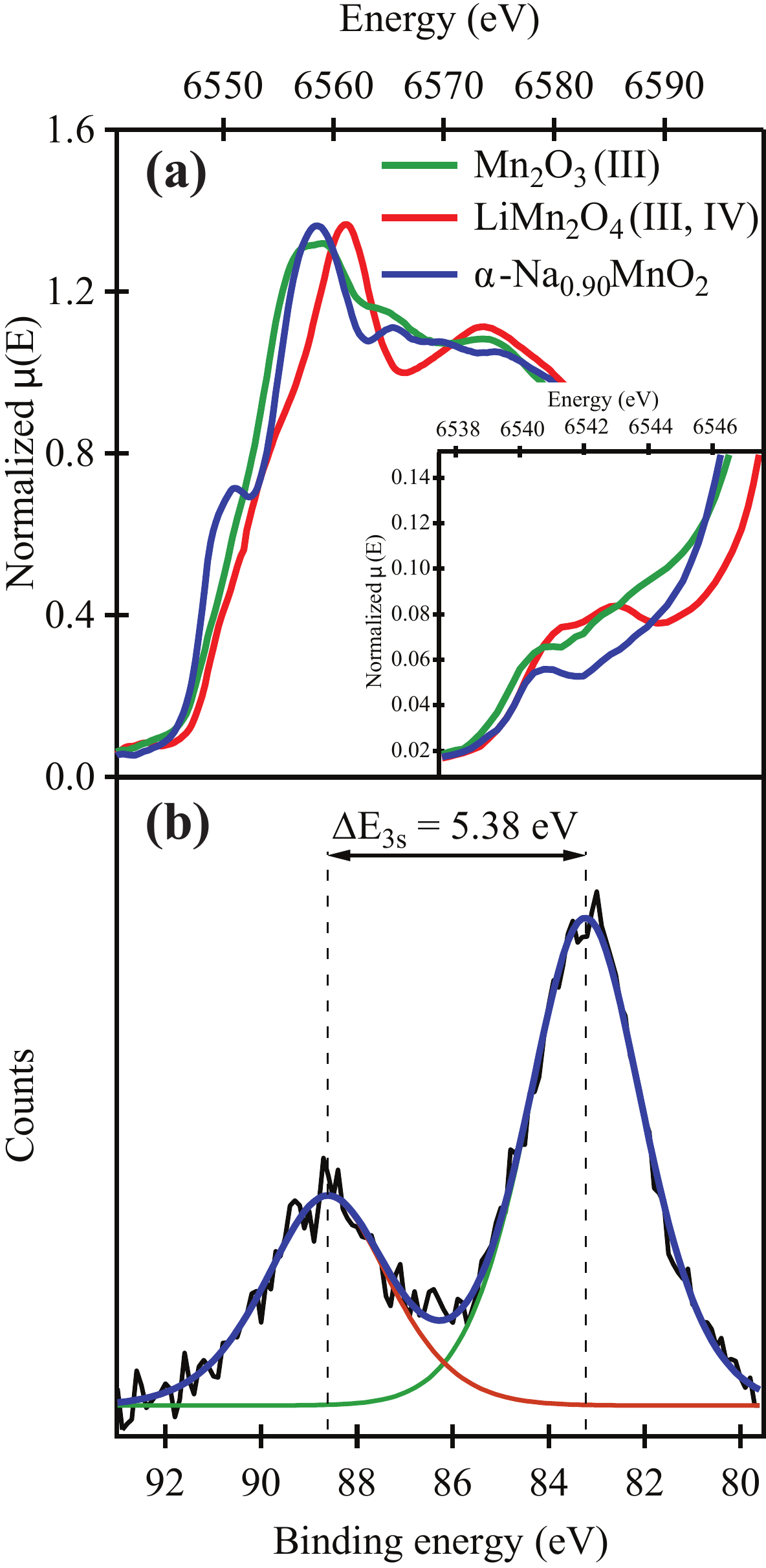}
\caption{(a) XANES data show the spectra collected for a 20 mm/hr grown ${\alpha}$-Na$_{0.90}$MnO$_2$ sample and standards at the Mn K-edge, and the inset details the pre-edge region of the spectra. (b) XPS data collected on a ${\alpha}$-Na$_{0.90}$MnO$_2$ sample showing the multiplet splitting of the Mn 3$s$ peak, which exhibits a $\Delta E=5.38$ eV, corresponding to an average valence of Mn$^{3.07+}$. }
\end{figure}

\begin{figure}[!p]
\includegraphics[scale=0.32]{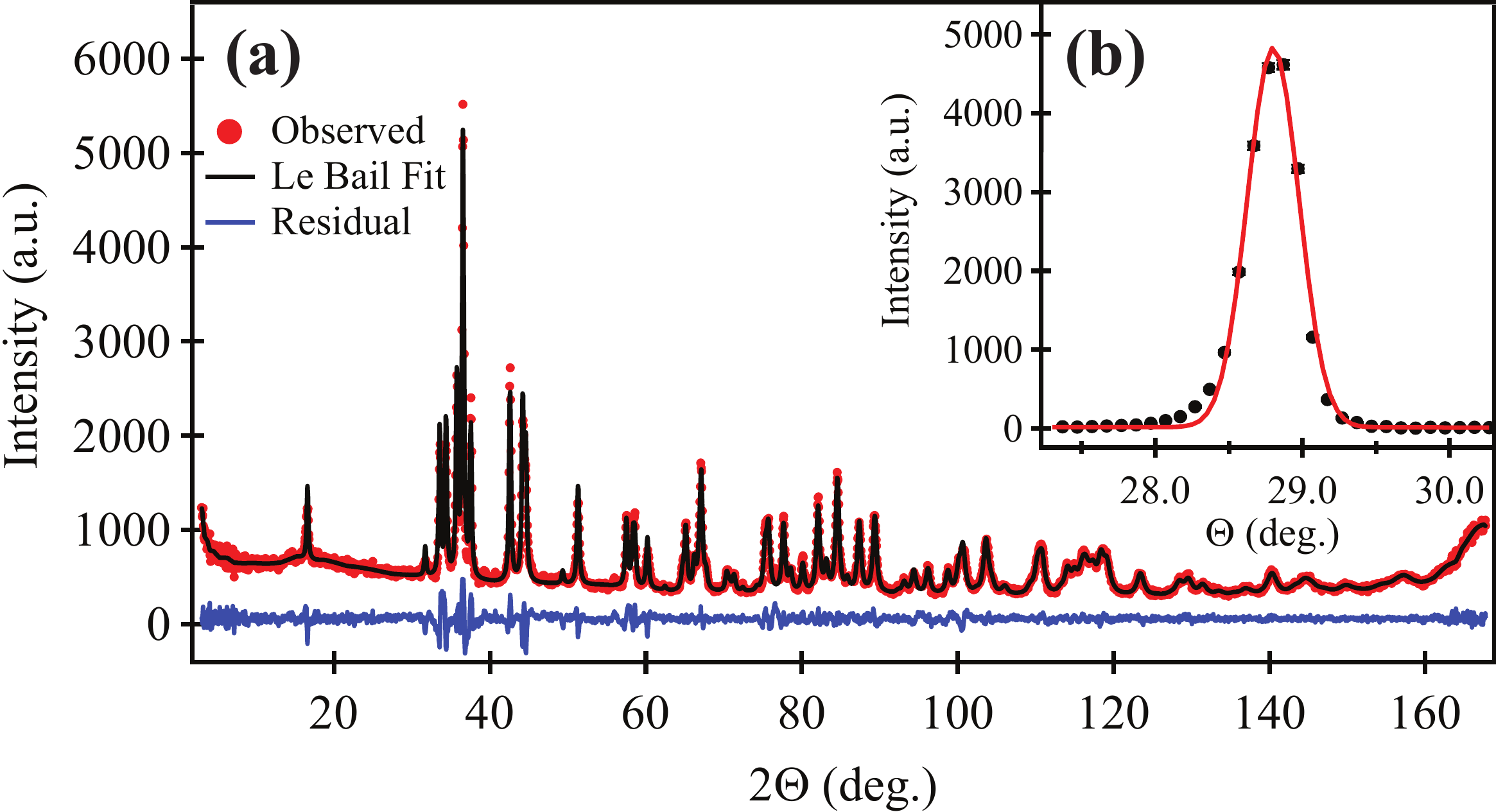}
\caption{a) Neutron powder diffraction data and corresponding Le Bail refinement for a crushed single crystal of $\alpha$-Na$_{0.90}$MnO$_2$ (b) Bulk averaged mosaic of a crystal shown through the rocking curve collected at the (200) nuclear Bragg peak of a typical crystal of $\alpha$-Na$_{0.90}$MnO$_2$.  Solid line shows a Gaussian fit to the peak of the form $I\propto exp(-\frac{1}{2}{(\frac{x-x_0}{w})}^2$) and where the $FWHM=2w\sqrt{2ln(2)}$ defines the mosaic, which after taking into account the instrumental resolution defines the intrinsic mosaic spread of the crystal.}
\end{figure}

\begin{figure}[!p]
\includegraphics[scale=0.445]{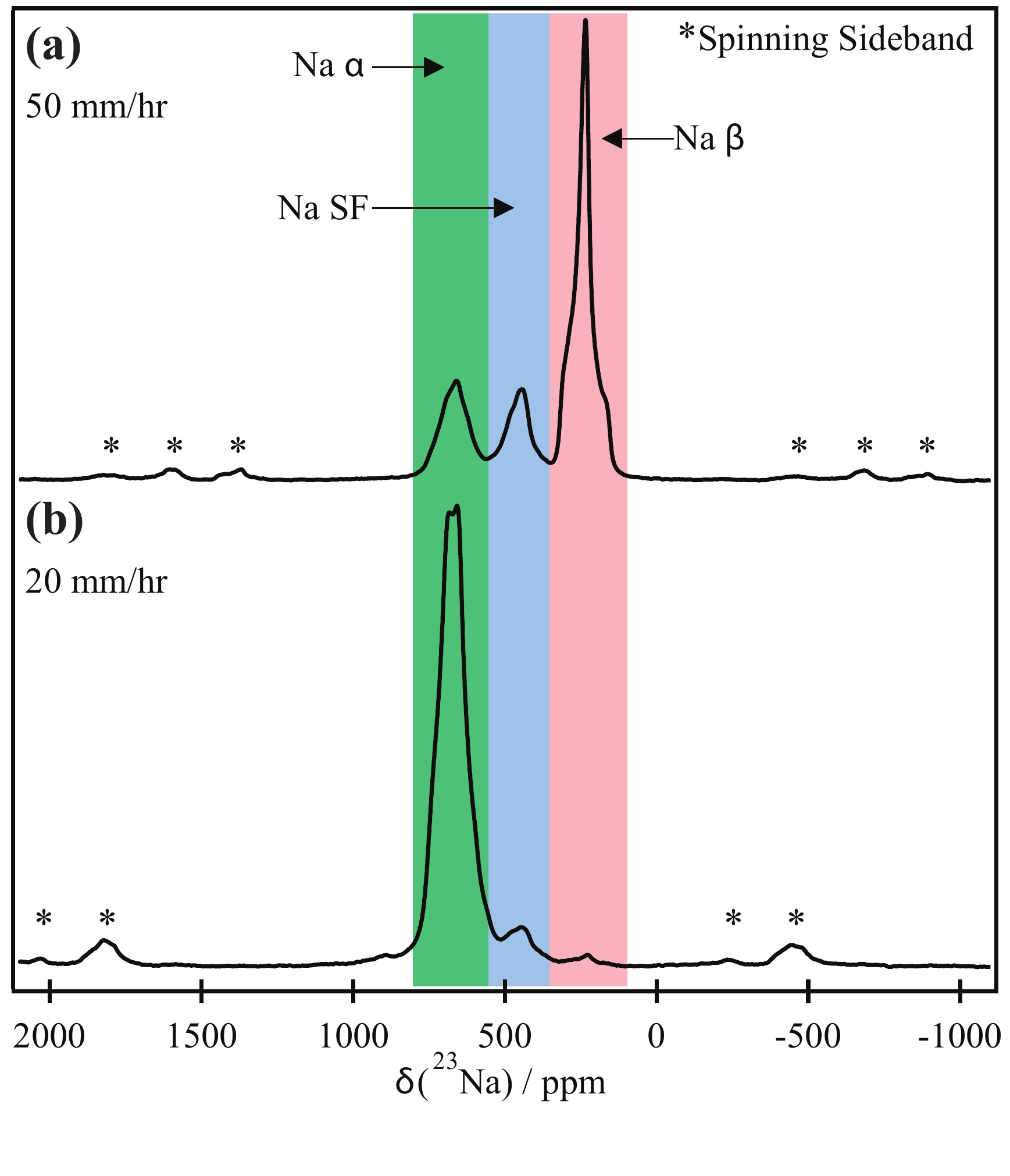}
\caption{$^{23}$Na ssNMR spectra obtained at room temperature at 4.7 T on two samples of single crystal NMO grown at different rates. The peaks corresponding to Na nuclei in $\alpha$-NaMnO$_2$ domains (Na $\alpha$), in $\beta$-NaMnO$_2$ domains (Na $\beta$), and in the vicinity of a stacking fault (Na SF), are shown on the figure. Spinning sidebands due to fast sample rotation are indicated by (*).}
\end{figure}

\end{document}